\documentclass[manuscript]{aastex}
\shorttitle{In-Situ Particle Acceleration  and the Problem of
Background Plasma Overheating} \shortauthors{Chernyshov et al.}

\begin{document}

\title{Stochastic Particle Acceleration and the Problem of
Background Plasma Overheating}

\author{D.O.Chernyshov\altaffilmark{1,2}, V.A. Dogiel\altaffilmark{1,2}, C.M. Ko\altaffilmark{2,3}}
\affil{$^1$I.E.Tamm Theoretical Physics Division of P.N.Lebedev Institute, Leninskii
pr, 53, 119991 Moscow, Russia}
\affil{$^2$Institute of Astronomy, National Central University, JhongLi 320, Taiwan}
\affil{$^3$Deparmtent of Physics and Center for Complex Systems, National Central University, JhongLi 320, Taiwan}
\email{cmko@astro.ncu.edu.tw}

\begin{abstract}
The origin of hard X-ray (HXR) excess emission from clusters of
galaxies is still an enigma, whose nature  is debated. One of the
possible mechanism to produce this emission  is the bremsstrahlung
model. However, previous analytical and numerical calculations
showed that in this case the intracluster plasma had to be
overheated very fast because suprathermal electrons emitting the
HXR excess lose their energy mainly by Coulomb losses, i.e., they
 heat the background plasma. It was concluded also from
these investigations that it is problematic to produce emitting
electrons from a background plasma by stochastic (Fermi)
acceleration because the energy supplied by external sources in
the form of Fermi acceleration is quickly absorbed by the
background plasma. In other words the Fermi acceleration is
ineffective for particle acceleration. We revisited this problem
and found that at some parameter of acceleration the rate of
plasma heating is rather low and the acceleration tails of
non-thermal particles can be generated and exist for a long time
while the plasma temperature is almost constant. We showed also
that for some regime of acceleration the plasma cools down instead
of being heated up, even though external sources (in the form of
external acceleration) supply energy to the system. The reason is
that the acceleration withdraws effectively high energy particles
from the thermal pool (analogue of Maxwell demon).
\end{abstract}


\maketitle

\section{Introduction}\label{intro}

One of the most important problem in astrophysics is the problem
of particle acceleration. The general expression for acceleration
of a charged particle is
\begin{equation}
  \frac{d}{dt}(\gamma m{\bf v})=Ze\left({\bf E}+\frac{1}{c}{\bf v}\times {\bf B}\right)\,,
\end{equation}
where ${\bf E}$ and ${\bf B}$ are the electric and magnetic field strength,
${\bf v}$ is the velocity of particle and $\gamma = 1/\sqrt{1 - v^2/c^2}$.
In most astrophysical conditions static electrical fields cannot be maintained because of
a very high electrical conductivity. Therefore the acceleration can be
associated either with non-stationary electric fields
(electromagnetic waves) or with time-varying magnetic fields.
In the latter case the work can be done by the induced electric field
\begin{equation}
  \frac{1}{c}\frac{\partial {\bf B}}{\partial t}=-{\bf\nabla\times E}\,.
\end{equation}
The basic idea of acceleration by electromagnetic inhomogeneities
in astrophysical conditions was suggested by \citet{fermi1,fermi2}
who assumed that the Galactic cosmic rays (CRs) were accelerated by collisions
of charged particles with fluctuations of magnetic fields (magnetic clouds) moving
chaotically with the velocity dispersion $u$.
One of the features of this theory was that it yielded naturally a power-law spectrum of
accelerated particles. The rate of particle acceleration by this stochastic mechanism is
about
\begin{equation}
  \left(\frac{d\mathcal{E}}{dt}\right)_F\sim \frac{u^2}{v^2\tau}\mathcal{E}\,,
\end{equation}
where $\mathcal{E}$ and $v$ are the particle kinetic energy and velocity,
and $\tau$ is the average time of  particle collision with the clouds.
This rate of acceleration is slow because  $u\ll v$.
In kinetic equations, the Fermi (stochastic) acceleration is described as momentum diffusion
\citep[see e.g.,][]{topt85}
\begin{equation}\label{toptyginEq}
  \frac{\partial f}{\partial t}-\frac{1}{p^2}\frac{\partial}{\partial p}
  \left[D_F(p)p^2\frac{\partial f}{\partial p}\right]+\hat{L}f=0\,,
\end{equation}
with the diffusion coefficient $D_F(p)$ in the form
\begin{equation}\label{dfermi}
  D_F(p)\sim p^2\frac{u^2}{v^2\tau} \,.
\end{equation}
Here $f(p,t)$ is the particle distribution function, $p$ is the
particle momentum, $t$ is the time, and the operator $\hat{L}$ describes particle
spatial propagation and their momentum losses.

In spite of its low efficiency, stochastic acceleration may be
essential for particle acceleration in solar flares \citep[see e.g.,][]{miller,petr12},
in the interstellar medium of the Galaxy
\citep[][]{ber90} and near the Galactic center \citep[see e.g.,][]{mertsch}.

The problem of stochastic particle acceleration in galaxy clusters
arose  from  observations in the hard X-ray (HXR) energy range
\citep[see e.g.,][]{fusf,fus,repha,reph,eckert,neva09,ajello1}
which showed an emission excess above the equilibrium thermal
X-ray spectrum.

One of the several interpretations of the HXR excess from the Coma cluster
in the range 20-80 keV was an assumption that it was produced by bremsstrahlung
radiation of suprathermal electrons \citep[see e.g.,][]{elb99} accelerated in
the intracluster medium.
However, this model was criticized by \citet{petr01} who concluded
from simple estimates that in this case the intracluster plasma in
Coma had to be overheated very fast.
The point is that suprathermal electrons emitting the HXR excess lose their energy
mainly by Coulomb losses, i.e., they lose their energy by heating the background plasma.
If these electrons generate an X-ray flux $L_X$ by bremsstrahlung,
they transfer the energy flux $L_C$ to the background plasma.
The necessary energy input is estimated as
\begin{equation}\label{c_en}
  L_C\sim L_X\left[\frac{(d\mathcal{E}/dt)_C}{(d\mathcal{E}/dt)_{BR}}\right]\,,
\end{equation}
where $(d\mathcal{E}/dt)_C$ and $(d\mathcal{E}/dt)_{BR}$ are the
rates of Coulomb and bremsstrahlung losses, respectively.
In the keV energy range  $(d\mathcal{E}/dt)_C\gg (d\mathcal{E}/dt)_{BR}$,
and this seems to make plasma overheating inevitable.
However, we should point out that in effect particle acceleration may also be
accompanied by plasma cooling due to run-away flux of high energy
particles from thermal pool, and more careful analysis is necessary
to define which of these effects (plasma heating or cooling) prevails.
This analysis is presented in the following sections.

\section{ Review of Particle Acceleration from Background Plasma}\label{lin}

A natural source for suprathermal particles is stochastic acceleration
of seed particles from a background plasma.
These particles are accelerated when the rate of acceleration
$(d\mathcal{E}/dt)_F$ exceeds the rate of the Coulomb losses $(d\mathcal{E}/dt)_C$.
A characteristic energy called the injection energy $\mathcal{E}_{\rm inj}$ is the
energy above which a non-thermal spectrum is formed by acceleration.
It is determined by equating these rates of acceleration and loss.

The kinetic equation in the particle momentum space
describing stochastic particle acceleration from background plasma
has the form (assume isotropic distribution)
\begin{equation}\label{e_k}
  {{\partial f}\over{\partial t}}+{1\over p^2}{\partial\over{\partial p}}p^2
  \left[\left(\frac{dp}{dt}\right)_C f - \left\{D_C(p)+D_F(p)\right\}{{\partial f}
  \over{\partial p}}\right]=0\,,
\end{equation}
where $D_F(p)$ is the diffusion coefficient of stochastic (Fermi)
acceleration, and $(dp/dt)_C$ and $D_C(p)$ describe particle
momentum losses and diffusion due to Coulomb collisions. These
coefficients are calculated from the total distribution function
$f$ (see Appendix~\ref{appendix-kin}), and therefore in general
the equation is nonlinear.

\subsection{Linear Approximation}

An analytical solution of this equation for the case of weak
acceleration from a background plasma with temperature $T$ was
obtained by \citet{gur60}. The term of stochastic acceleration was
taken in the phenomenological form
\begin{equation}
  D_F(p)=\alpha p^2\,.
\end{equation}
The analysis was provided for the case when the characteristic
time of stochastic acceleration,
\begin{equation}\label{tauF}
  \tau_F=p^2/D_F\,,
\end{equation}
is much larger than the time of thermal particle collisions,
$\tau_{th}$,
\begin{equation}\label{tauth}
  \tau_{th}\simeq \sqrt{\frac{2}{m}}~\frac{m_e(k_{\rm B}T)^{3/2}}{\pi N e^4\ln\Lambda}\,,
\end{equation}
where $N$ is the density and $T$ is the temperature of background
plasma, $\ln\Lambda$ is the Coulomb logarithm, $m_e$ is the
electron rest mass and $m$ is the mass of accelerated particles.

\begin{figure}[ht]
\begin{center}
\includegraphics[width=0.6\textwidth]{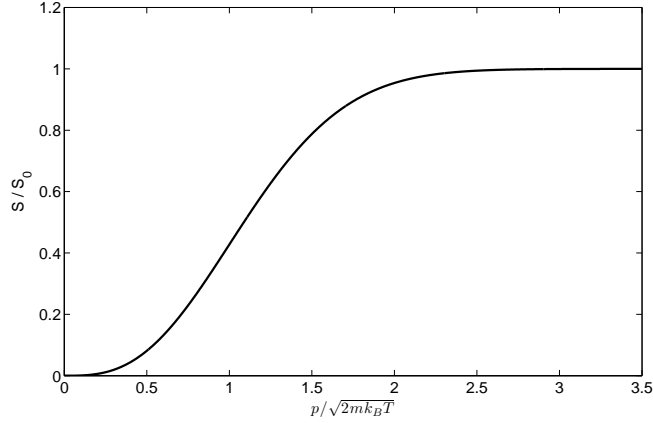}
\end{center}
\caption{Run-away flux $S$ as a function of the dimensionless
momentum $p/\sqrt{2mk_{\rm B}T}$.}
\label{fig:S0}
\end{figure}

In this case the injection energy $\mathcal{E}_{\rm inj}$ is much
larger than the plasma temperature, $\mathcal{E}_{\rm inj}\gg k_{\rm
B}T$. Coulomb collisions keep the equilibrium Maxwellian
distribution for most part of the momentum range, and the coefficients
of Eq.~(\ref{e_k}) for nonrelativistic momenta $p\gg\sqrt{2mk_{\rm B}T}$
\citep[as used by][]{gur60} for the Maxwellian distribution
function. For $\tau_{th}\ll\tau_F$, significant distortions from
the equilibrium Maxwellian state are expected only for very large
values of momenta and a very small fraction of thermal particles
is accelerated. Therefore \citet{gur60} assumed that the number
of particles $N(t)$ in the momentum range $p<p_{\rm inj}$ varies
very slowly  with time $t$, $N(t)=N_0-St$, where $N_0$ is the
initial particle density and a small run-away flux $S$  is
generated at relatively high momentum range. The run-away
flux for the case of slow acceleration can be described as
\begin{eqnarray}\label{runaway}
  S(p)=S_0{4\over\sqrt{\pi}}\int\limits^{{\bar p}}_0
  x^2 e^{-x^2} dx
  =S_0\left[{\rm erf}\left({\bar p}\right)
  -{2\over\sqrt{\pi}}\,{\bar p}\,e^{-{\bar p}^2}\right]\,, %\nonumber
\end{eqnarray}
where ${\rm erf}(z)$ is the error function, ${\bar p}=p/\sqrt{2mk_{\rm B}T}$,
and the constant $S_0$ is derived from boundary conditions. The flux is zero
at $p=0$ but when $p \gg \sqrt{2mk_{\rm B}T}$, it reaches a maximum value
$S(p)=S_0$ as shown in Fig.~\ref{fig:S0}.

In the momentum range where $S(p)\simeq S_0$ the distribution function
is non-Maxwellian and it is described by the kinetic equation
\begin{equation}\label{eq_flux}
  p^2\left[\left(\frac{dp}{dt}\right)_C f -
  \left\{D_C(p)+D_F(p)\right\}{{\partial f}\over{\partial p}}\right]=S_0\,.
\end{equation}
The acceleration forms a nonthermal component of the spectrum in
the range $p>p_{\rm inj}$ where $p_{\rm inj}$ is the solution of equation
\begin{equation}\label{pinj}
  p_{\rm inj} = {D_F(p_{\rm inj})\over (dp/dt)_C}\,.
\end{equation}
However, if $D_F(p)\neq 0$ in the range $p<p_{\rm inj}$, a solution of this equation
describes also an excess of the distribution function above the equilibrium
 Maxwellian distribution
in momentum ranges both above and below $p_{\rm inj}$
\citep[see][]{gur60}. This excess at $p<p_{\rm inj}$ is formed by
Coulomb collisions in the transition range  between the thermal
(Maxwellian) and non-thermal parts of the spectrum.  If the HXR
excess is due to bremsstrahlung emission of electrons from this
transition region then the relation~(\ref{c_en}) used by
\citet{petr01} cannot be applied to the estimate of $L_C$ and more
accurate calculations are necessary.

Bremsstrahlung emission of electrons from the transition region
was calculated in \citet{dog1,liang,dog2}. The conclusion is that
the necessary energy input  $L_C$ for Coma was about one order of
magnitude less than obtained by \citet{petr01}. This may solve the
problem of the plasma overheating. However, their linear analysis
of equation~(\ref{e_k}) does not include variations of temperature
$T$ which is supposed to be constant.

\subsection{Non-Linear Treatment}

More reliable conclusions can be derived from analyses of the
nonlinear equation in the form similar to those used by
\citet{rosen}, when a feedback of accelerated particles on the
plasma temperature is taken into account. Very recently
\citet{wolfe} and \citet{east} provided similar numerical analysis
for the case of stochastic acceleration from a background plasma.

The nonlinear kinetic equation describing particle Coulomb
collisions is derived in \citet{ll} (see Appendix
\ref{appendix-kin}). Using this theory \citet{nayak} derived
coefficients of this equation for the case of  isotropic and
homogeneous distribution function for non-relativistic and
ultra-relativistic particles. Later, \citet{wolfe} extended their
analysis to the general case of anisotropic distribution function.

Numerical analysis of these equations has been performed by
\citet{wolfe} for the isotropic stochastic acceleration in the form
\begin{equation}
  D_F(p) = \alpha p^\varsigma \theta(p - 1/2)\,.
\end{equation}
\citet{wolfe} stated that
the continuous stochastic acceleration of thermal electrons
produced a nonthermal tail.
But for the hard X-ray emission in the Coma Cluster this model
actually cannot work because the energy gained by the particles is
distributed to the whole plasma on a timescale much shorter than
that of the acceleration process itself. Moreover, bremsstrahlung is
relatively inefficient to cool the accelerated electrons,
the energy of this tail is quickly dumped into the thermal background plasma and
heat the plasma.

Similarly, \citet{east} obtained numerical solutions of the nonlinear
isotropic kinetic equations  which included effects of plasma
heating for the stochastic diffusion in the form
\begin{equation}\label{petrosian}
  D_F(\mathcal{E})=\frac{\mathcal{E}^2}
  {\zeta(\mathcal{E})\tau_0(1+\mathcal{E}_c/\mathcal{E})^q}\,,
\end{equation}
where
$\mathcal{E}=\sqrt{p^2+1}-1$ is the kinetic energy normalized to $mc^2$,
$\zeta(\mathcal{E})=(2-\gamma^{-2})/(1+\gamma^{-1})$, and
$\tau_0$, $\mathcal{E}_c$ and $q$ are free parameters.

\citet{east} concluded that their calculations confirmed
qualitatively results of \citet{dog2} that the required input
energy $L_C$ was lower than that follows from the estimate (\ref{c_en})
but by a factor of 2 or 3 only, and that did not solve the problem of
plasma overheating. Besides, they argued that their calculations
confirmed results of \citet{wolfe} that stochastic acceleration
could not work in clusters because the energy gained by the
particles was distributed to the whole plasma on timescales much
shorter than that of the acceleration process. At acceleration
rates smaller than the thermalization rate of the background
plasma, there is very little acceleration. The primary effect of
acceleration is heating of the plasma. In the opposite case, at
higher energizing rates, a distinguishable nonthermal tail is
developed, but this is again accompanied by an unacceptably high
rate of heating.

In other words it follows from these investigations that it is
problematic to accelerate particles from  a background plasma
because the main effect of this acceleration is plasma
overheating. The energy supplied by external sources in the form
of stochastic (Fermi) acceleration is quickly absorbed by a background plasma.
An interesting question arises: whether any
conditions exist when the stochastic acceleration generate
prominent nonthermal tails while the plasma is not overheated and
its temperature varies relatively slowly.
From the analysis in the following sections, we argue that the answer is affirmative.

\section{Particle Acceleration from Background Plasma: Quasi-Linear
Approximation}\label{quasilin}

First, we estimate variations of plasma temperature derived in
quasi-stationary approximations when the distribution function can
be presented as $f = f(p,N,T)$. In this case,
\begin{equation}\label{q_st}
  \frac{\partial f}{\partial t} = \frac{\partial f}{\partial N}\frac{dN}{dt}
  + \frac{\partial f}{\partial T}\frac{dT}{dt}\,.
\end{equation}
where $N =N(t)$ and $T = T(t)$ are slowly varying functions of $t$.

\subsection{Distribution function}\label{distrib_f}

In this subsection we investigate the isotropic form of the kinetic equation~(\ref{equp}).
This equation describes stochastic particle acceleration from background plasma and
it is exactly the same as Eq.~(\ref{e_k}). The appropriate boundary conditions
are Eqs.~(\ref{kinet_symp}) \& (\ref{kinet_sym1p}).
Recall that the particle momentum has been normalized to $mc$.
Here and in the following the temperature $T$ is indeed the thermal energy $k_{\rm B}T$
normalized to $mc^2$. The particle kinetic energy $\mathcal{E}=\sqrt{p^2+1}-1$
is also normalized to $mc^2$. The coefficients $(dp/dt)_C(p)$, $D_C(p)$ and $D_F(p)$ are
normalized accordingly.

The stochastic Fermi acceleration is supposed to be isotropic and has a
phenomenological form as
\begin{equation}\label{pcut}
  D_F(p) = \alpha p^\varsigma \theta(p-p_0)\,,
\end{equation}
where $\alpha$, $\varsigma$ and $p_0$ are arbitrary parameters.
The problem is characterized also by the injection momentum
\begin{equation}\label{pinjection}
  \alpha p_{\rm inj}^\varsigma
  = - p_{\rm inj} \left.\left({dp\over dt}\right)_c\right|_{p=p_{\rm inj}}\,.
\end{equation}
The acceleration is effective in the momentum range
$p>{\rm max}\{p_0,p_{\rm inj}\}$.

Similar to \citet{gur60} we assume that the acceleration time $\tau_F$,
is much longer than the time of thermal particle collisions $\tau_{th}$,
i.e., values of $p_{\rm inj}$ or $p_0$ are large and
one of the corresponding energy values  is much higher
than the temperature,
\begin{equation}
  T\ll {\rm max}(\mathcal{E}_{\rm inj}, \mathcal{E}_0)\,.
\end{equation}
In this case, Coulomb collisions keep the equilibrium Maxwellian
distribution over an extended momentum range with a significant
deviation from this distribution at very large momenta,
i.e., a small part of thermal particles is accelerated. The number
of non-thermal particles generated by the acceleration $N_n$ in
this case is  much smaller than the number of thermal particles
$N$, $N_n/N\ll 1$.

Below we present the distribution function and the coefficients
of the kinetic equation as series expansions over the small
parameter $\epsilon=N_n/N\ll 1$,
\begin{eqnarray}
  f(p,t) &&= f_0(p,t) + f_1(p,t) + O\left(\epsilon^2\right)\,, \nonumber \\
  D_c(p,t) &&= D_0(p,t) + D_1(p,t) + O\left(\epsilon^2\right)\,,  \\
  \left({dp\over dt}\right)_c(p,t) &&= \left({dp\over dt}\right)_0(p,t)
  + \left({dp\over dt}\right)_1(p,t) + O\left(\epsilon^2\right)\,. \nonumber
\end{eqnarray}
Here $O\left(\epsilon^i\right)$ denotes terms of order $\epsilon^i$ or above.
Note that $f_i(p,t)=O(\epsilon^i)$,
$D_0$ and $(dp/dt)_0$ are calculated from Eq.~(\ref{coef1p}) for the function $f_0$,
and $D_1$ and $(dp/dt)_1$ for the function $f_1$, etc.

In the quasi-stationary approximation the derivative $\partial f/\partial t$ can be
presented in the form (\ref{q_st}). The derivatives $dN/dt$ and
$dT/dt$ can be presented as series $dN/dt = O(\epsilon)$ and
$dT/dt = O(\epsilon)$, because without acceleration ($N_n=0$) we have
$dN/dt = 0$ and $dT/dt = 0$. Here we have
\begin{equation}
  \frac{\partial f}{\partial t} = \frac{\partial f_0}{\partial t} +
  O\left(\epsilon^2\right)\,,
\end{equation}
and ${\partial f_0}/{\partial t}$ is of the order of $\epsilon$.

It is convenient to express the distribution function as
\begin{equation}\label{ff}
  f(p) = f^I(p)\theta(p_0 - p) + f^{II}(p)\theta(p-p_0)\,.
\end{equation}

First, we find the solution of Eq.~(\ref{e_k}) in the momentum
range $0<p<p_0$ where the acceleration term vanishes and $f=f^I$
(see Eq.~(\ref{ff})). In zero order of expansion  (no
acceleration) the function $f_0$ is  Maxwellian
\begin{equation}\label{maxwellI}
  f^I_0(p)
  =C_0 \exp\left[\int_0^p\left({dp\over dt}\right)_0{dp\over D_0}\right]
  =C_0\exp(-\mathcal{E}/T)\,,
\end{equation}
where $D_0$ and $(dp/dt)_0$ are the Maxwellian kinetic coefficients.
For $p \gg \sqrt{T^2+1}-1$ the Bethe-Bloch approximation for the
these coefficients is
\begin{eqnarray}
  \left({dp\over dt}\right)_0 &&= - A \left(1+{1\over p^2}\right)\,, \label{bethe-1} \\
  D_0 &&= - T\sqrt{1+{1\over p^2}}\left({dp\over dt}\right)_0
  =AT\left(1+{1\over p^2}\right)^{3/2}\nonumber \,.
\end{eqnarray}
Here and below
\begin{equation}\label{taumc-1}
  A = 4\pi r_e^2cN \ln \Lambda\,.
\end{equation}

The characteristic time of Coulomb losses for a particle with
momentum $p$ (in unit of $mc$) is
\begin{equation}\label{tauC}
  \tau_C(p)\sim \frac{p^3}{A(p^2+1)}\,.
\end{equation}
Constant $C_0$ is estimated from the normalization condition
\begin{equation}\label{maxwellC0}
  C_0 = N \left[\int\limits_0^{p_0} p^2 f^I_0(p)dp\right]^{-1} \approx
  \frac{N\exp(-T^{-1})}{TK_2(T^{-1})}\,,
\end{equation}
where $K_2(x)$ is the modified Bessel function.
For non-relativistic temperatures ($T\ll 1$) we obtain
\begin{equation}
  C_0 \approx N\sqrt{\frac{2}{\pi}}T^{-3/2}\,.
\end{equation}

The kinetic equation for the function $f^I_1$ can be rewritten as
\begin{equation}\label{kin-o2-epsilson}
  \frac{1}{p^2}\frac{\partial}{\partial p}p^2\left[
  D_0(p)\frac{\partial f^I_1}{\partial p} +
  D_1(p)\frac{\partial f^I_0}{\partial p} -
  \left(\frac{dp}{dt}\right)_0 f^I_1 -
  \left(\frac{dp}{dt}\right)_1 f^I_0\right] =
  \frac{\partial f^I_0}{\partial t} + O\left(\epsilon^2\right)\,.
\end{equation}
Integrating the above equation from 0 to $p$ gives
\begin{equation}\label{eq_nr_I_1-1}
  p^2 \left[ D_0(p)\frac{\partial f^I_1}{\partial p} -
  \left(\frac{dp}{dt}\right)_0 f^I_1\right] = -S =
  -\left(S_1 + S_2\right)\,,
\end{equation}
where $S$ is the flux of particles through the point $p$. Here
\begin{equation}\label{s1-1}
  S_1 = -\frac{dN(p,t)}{dt} = -\frac{\partial}{\partial t}\int\limits_0^p u^2f^I_0(u)du\,,
\end{equation}
\begin{equation}
  S_2 =  p^2\left[D_1(p)\frac{\partial f^I_0}{\partial p} -
 \left({dp\over dt}\right)_1 f^I_0(p)\right]\,.
\end{equation}
The flux $S_1$ describes a particle leakage (in momentum space) caused by the
acceleration. It generates a slow decrease of particle number in
the thermal region. The flux $S_2$ causes the plasma heating and
temperature variations with time.

Thus, the solution of Eq.~(\ref{eq_nr_I_1-1}) is
\begin{equation}\label{sol_eq_nr_I_1-1}
  f^I_1(p) = \exp(-\mathcal{E}/T)\left[C_1 - \int\limits_0^p
  \frac{S(u)}{u^2 D_0(u)}\exp(\mathcal{E}/T)du\right]\,.
\end{equation}
As in \citet{gur60} the value of the constant $C_1$ can be derived
from the normalization condition
\begin{equation}\label{C1_norm}
  \int\limits_0^{p_0} p^2 f^I_1(p)dp = 0\,.
\end{equation}
The kinetic coefficients $D_1$ and $(dp/dt)_1$ are calculated for
the function $f^I_1+f^{II}$. Therefore Eq.~(\ref{sol_eq_nr_I_1-1})
is an integral equation for $f^I_1(p)$ which
should be added by an equation for $f^{II}(p)$. The asymptotic
form of $f^I_1(p)$ for large values of $p$ can easily be derived.
Indeed, if $\mathcal{E}\gg T$ then $S_1(p) = O(\epsilon)$ while
$S_2(p) \sim O(\epsilon)\exp(-\mathcal{E}/T) \ll S_1$. Therefore
$S_2(p)$ can be neglected. As one can see from Eq.~(\ref{s1-1}) the
flux $S_1(p)$ remains almost constant for sufficiently large $p$
(see Fig.~\ref{fig:S0}). So with a high degree of accuracy we can put
$S_1(p) = S_N \equiv - dN/dt$, which is the same as
$S_0$ in \citet{gur60}.

It follows from Eqs. (\ref{sol_eq_nr_I_1-1}) and (\ref{C1_norm}) that the
constant $C_1$ is
\begin{equation}
  C_1 \approx S_N\tau_C(p_0)\sqrt{\frac{2}{\pi}}T^{-3/2}\,,
\end{equation}
where $\tau_C(p_0)$ is the characteristic time of Coulomb
collision for the particle momentum $p=p_0$ (for $\tau_F$ and
$\tau_C$ see Eqs. (\ref{tauF}) and (\ref{tauC})). Thus, for the
estimation of $S_N$ obtained in subsection~\ref{trans} we have
\begin{equation}
  C_1 \sim {N\over T^{3/2}}
  \exp\left(-\frac{\mathcal{E}_0}{T}\right)\frac{\tau_C(p_0)}{\tau_F(p_0)}
  \sim {N\over T^{3/2}}\exp\left(-\frac{\mathcal{E}_0}{T}\right)\ll C_0\,.
\end{equation}
The distribution function in the range $p<p_0$ can be written as
\begin{eqnarray}\label{rel_f-1}
  &&f^I(p)\simeq f^I_0(p)+f^I_1(p) \nonumber \\
  &&\quad\quad
  =\frac{N}{TK_2(T^{-1})}\exp\left(-\frac{\xi}{T}\right)
  -\frac{S_N}{AT}\left[\frac{1}{T}
  \exp\left(-\frac{\xi}{T}\right)Ei\left(\frac{\xi}{T}\right)
  - \frac{1}{\xi}\right] \,,
\end{eqnarray}
where $\xi = \sqrt{p^2+1} = \mathcal{E}+1$ is the total energy of particle and
\begin{equation}
  Ei(z) = \int\limits_{-\infty}^z\frac{\exp(x)}{x}dx\,.
\end{equation}
For non-relativistic  temperatures $\xi/T \gg 1$ the  expansion of $Ei(z)$ for
$z\gg 1$ is
\begin{equation}
  Ei(z) = \frac{\exp(z)}{z}\sum\limits_{k=0}^\infty \frac{k!}{z^k}\,.
\end{equation}
Thus for large values of $p$
\begin{equation}\label{fI-1}
  f^I(p) =
  \sqrt{\frac{2}{\pi}}\frac{N}{T^{3/2}}\exp\left(-\frac{\mathcal{E}}{T}\right)
  - \frac{S_N}{A(p^2+1)}\,.
\end{equation}
The distributions function Eq.~(\ref{rel_f-1}) can be presented in the form
\begin{equation}\label{f_1_bigO-1}
  f^I(p) = \left \{
  \begin{array}{l}
  f_0^I(p)
  + O(\epsilon)\,,\mbox{ for }\mathcal{E}\leq T   \\
  f_0^I(p)
  - \frac{S_N}{A(p^2+1)}+O\left(\epsilon^2\right)\,,
  \mbox{ for }\mathcal{E}_0\geq\mathcal{E}\gg T
  \end{array}
  \right.
\end{equation}
In the range $p \geq p_0$ the acceleration cannot be neglected.
With the constant flux $S_N$ of particles the equation for the
distribution function $f^{II}$ in this region reads
\begin{equation}\label{eq+acc-1}
  p^2\left[ \left\{D_0(p)+D_F(p)\right\}\frac{\partial f^{II}}{\partial p}
  -\left(\frac{dp}{dt}\right)_0 f^{II}\right] = -S_N\,.
\end{equation}
The general solution of this equation is \citep[see, e.g.,][]{gur60}
\begin{eqnarray}\label{gur_sol01-1}
  &&f^{II}(p) = C^{II} \exp\left\{\int \limits_0^p
  \frac{(dp/dt)_0(u)du}{D_F(u)+D_0(u)}\right\} \nonumber \\
  &&- S_N \exp\left\{\int \limits_0^p \frac{(dp/dt)_0(u)du}{D_F(u)+D_0(u)}\right\}
  \int\limits_0^p\frac{v^{-2}dv}{D_F(v)+D_0(v)}
  \exp\left\{-\int\limits_0^v \frac{(dp/dt)_0(u)du}{D_F(u)+D_0(u)}\right\}\,.
\end{eqnarray}
The constant $C^{II}$ can be estimated from the continuity
condition  at $p=p_0$: $f^I(p_0) = f^{II}(p_0)$ while the value of
$S_N$ can be estimated from the second boundary condition:
$f^{II}(p_{\rm max}) = 0$.

For $p \gg p_{\rm inj}$, we can assume acceleration dominates Coulomb loss. It is
easy to show from Eq.~(\ref{eq+acc-1}) and Eq.~(\ref{pcut}) that the function
$f^{II}(p)$ is a power-law
\begin{equation}\label{fII_pwlaw-1}
  f^{II}(p) = \tilde{C}_1 + \frac{S_N}{\alpha(\varsigma+1)}p^{-\varsigma-1}\,,
\end{equation}
where $\tilde{C}_1$ is a constant.

\subsection{Plasma heating rate}\label{PHR}

Using the total distribution function $f$ (see Eq.~(\ref{ff}), where $f^I$ and
$f^{II}$ are determined by Eqs.~(\ref{f_1_bigO-1}) \& (\ref{gur_sol01-1})),
we can calculate the kinetic coefficients~(\ref{coef1p}) for the nonlinear equation~(\ref{e_k}),
and then estimate the temperature variations of the
background plasma caused by particle acceleration.
In this case the stochastic Fermi momentum diffusion describes the energy supply into the system
by external sources.
Generally speaking, energy supply can vary with time, but usually it is assumed that
external sources keep a
stationary level of acceleration such that $D_F$ is constant.

The total energy input into the system is
\begin{equation}
  \dot{W}_{\rm ext} =-\int\limits_0^{\infty} \mathcal{E}\frac{\partial}
  {\partial p}\left[p^2D_F\frac{\partial f}{\partial p}\right]\, dp\,.
\end{equation}
It is a function of time even if $D_F$ is constant,
because the distributions function $f$ is time dependent.

Note that Coulomb collisions do not change the total energy in the system,
therefore we have
\begin{equation}\label{cond}
  \int\limits_0^{\infty} \mathcal{E}\frac{\partial}{\partial p}
  p^2\left[\left(\frac{dp}{dt}\right)_C f -
 D_C(p){{\partial f}\over{\partial p}}\right]\, dp=0\,.
\end{equation}
This condition is valid for any function $f$ if the kinetic coefficients
$(dp/dt)_C$ and $D_C(p)$ are calculated from Eq.~(\ref{coef1p}) for this function $f$.

The energy supplied by the stochastic Fermi acceleration is distributed over
the spectrum in the form of accelerated particles and a heated plasma,
because accelerated particles lose their energy by Coulomb collisions
and thus transfer a part of their energy to thermal particles.
Variations of
$dT/dt$ in the quasi-equilibrium part of the spectrum can be
derived from estimates of the energy flux into the region $p<p_0$
which is
\begin{equation}
  \dot{W_0} = \frac{\partial }{\partial t}
  \int\limits_0^{p_0} p^2 \mathcal{E}f^I(p) dp \label{wdot_1}
  = \int\limits_0^{p_0} \mathcal{E}\frac{\partial}{\partial p}\left[p^2D_c
  \frac{\partial f^I}{\partial p} -p^2\left(\frac{dp}{dt}\right)_c f^I\right]\,dp\,,
\end{equation}
where the coefficients $D_c$ and $(dp/dt)_c$ are calculated for
the total distribution function (\ref{ff}). For the estimation of the integral
(\ref{wdot_1}) we can use the condition (\ref{cond}), and obtain
\begin{equation}
  \dot{W_0} = - \int\limits_{p_0}^\infty \mathcal{E}\frac{\partial}{\partial p}
  \left[p^2D_c\frac{\partial f^{II}}{\partial p} -
  p^2 \left(\frac{dp}{dt}\right)_c f^{II}\right]\,dp\,.
\end{equation}
Integration by parts gives
\begin{equation}\label{wdot_3_c-1}
  \dot{W_0}=-\mathcal{E}_0S_N + \int\limits_{p_0}^\infty
  \frac{p^3}{\sqrt{p^2+1}}\left[D_c\frac{\partial f^{II}}{\partial p}
  -\left(\frac{dp}{dt}\right)_c f^{II} \right]\,dp\,.
\end{equation}
Since $\dot{W_0} = O(\epsilon)$ and $f^{II} = O(\epsilon)$ we can use
the Maxwellian (Bethe-Bloch) expressions for the kinetic
coefficients $D_0$ and $(dp/dt)_0$ as in Eq.~(\ref{bethe-1}),
\begin{equation}\label{wdot_3-1}
  \dot{W_0}=-\mathcal{E}_0S_N + \int\limits_{p_0}^\infty
  \frac{p^3}{\sqrt{p^2+1}}\left[D_0\frac{\partial f^{II}}{\partial p}
  -\left(\frac{dp}{dt}\right)_0 f^{II} \right]\,dp\,.
\end{equation}

We see that the energy input into the thermal part of the spectrum
(plasma heating) is determined by two processes: (i) energy losses
of nonthermal particles (the integral of Eq.~(\ref{wdot_3-1})) which heat the
plasma, and (ii) a particle escape to the high energy part of the
equilibrium spectrum (the first term on the RHS of Eq.~(\ref{wdot_3-1})) which cools
the plasma. On the other hand we can express $\dot{W_0}$ in the form
\begin{equation}\label{w0_dot}
  \frac{dW_0}{dt} = \frac{\partial W_0}{\partial T}\frac{dT}{dt} +
  \frac{\partial W_0}{\partial N}\frac{dN}{dt}\,.
\end{equation}

\begin{figure}[ht]
\begin{center}
\includegraphics[width=0.6\textwidth]{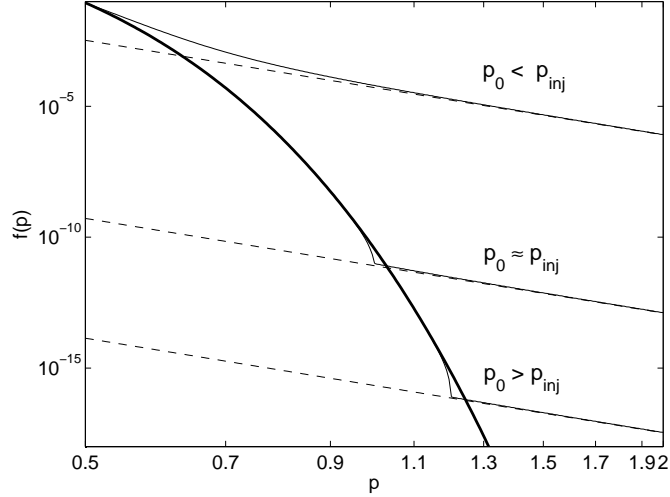}
\end{center}
\caption{Shape of $f(p)$ for different values of $p_0$. Thin solid
line represents $f(p)$, thick solid line - pure Maxwellian
distribution, dashed line - power-law approximation.}
\label{fig:spectr}
\end{figure}

To the first order of $\epsilon$ of the expansion of $dT/dt$ we
can take $W_0$ as
\begin{equation}\label{EnergyContent0}
  W_0 = \int\limits_0^{p_0}u^2 \mathcal{E} f_0^I(u)\,du =
  \frac{N\exp(-T^{-1})}{TK_2(T^{-1})}\int\limits_0^{p_0}u^2 \mathcal{E}
  \exp\left(-{\mathcal{E}\over T}\right) du \,. \nonumber
\end{equation}
In the general case the temperature variations can be calculated
numerically (see section~\ref{nl}). However, these calculations
can be simplified. The point is that the particle spectrum described by
Eqs.~(\ref{fI-1}) \& (\ref{gur_sol01-1}) depends strongly on the relation
between the momenta $p_{\rm inj}$ and $p_0$. Fig.~\ref{fig:spectr}
illustrates this situation: as $p_0$ increases the transition
region in momentum range $p>p_0$ shrinks and finally disappears
when $p_0$ reaches $p_{\rm inj}$. In the limiting case $p_0>p_{\rm inj}$
the transition region vanishes almost completely and the
power-law tail of nonthermal particles is attached almost directly to
the thermal equilibrium distribution. In this case,
evaluations of the plasma temperature can be performed
analytically because  the functions $f^I$ and $f^{II}$ have very
simple form. We notice that the conclusion of \citet{gur60} about a
very extended transition region between thermal and nonthermal
parts of the spectrum is valid only for the case when $p_0<p_{\rm inj}$.

\subsection{The case of transitionless acceleration}\label{trans}

If $p_0>p_{\rm inj}$, Eq.~(\ref{fII_pwlaw-1}) is an appropriate solution
for the distribution function.
$\tilde{C}_1$ and $S_N$ are determined from the boundary conditions at $p=p_0$ and
$p=p_{\rm max}$, namely, $f^{II}(p_0)=f^I(p_0)=f_0$ and $f^{II}(p_{\rm max})=0$,
\begin{eqnarray}
  S_N &&= \alpha (\varsigma+1)p_0^{\varsigma+1} f_0\,,\label{SN1-1} \\
  \tilde{C}_1 &&= -{S_Np^{-(\varsigma+1)}_{\rm max}\over\alpha(\varsigma+1)}
  =- f_0 \left({p_{\rm max} \over p_0}\right)^{-(\varsigma+1)}\,.
\end{eqnarray}
As $p_{\rm max} \gg p_0$, thus for simplicity we set $\tilde{C}_1 =0$,
and for non-relativistic temperatures  $T \ll 1$ from Eq.~(\ref{fI-1}) we have
\begin{equation}\label{rel_f0-1}
  f_0 = \sqrt{\frac{2}{\pi}}\frac{N}{T^{3/2}}\exp\left(-{\mathcal{E}_0\over T}\right)
  \left[1 +\frac{\alpha(\varsigma+1)p_0^{\varsigma+1}}{A(p_0^2+1)}\right]^{-1}\,.
\end{equation}
In this case the run-away particle flux toward high energies can
be expressed directly from Eq.~(\ref{SN1-1}) as
\begin{equation}\label{SN2-1}
  S_N = \alpha(\varsigma+1)p_0^{\varsigma+1}
  \sqrt{\frac{2}{\pi}}\frac{N}{T^{3/2}}\exp\left(-{\mathcal{E}_0\over T}\right)
  \left[1
  +\frac{\alpha(\varsigma+1)p_0^{\varsigma+1}}{A(p_0^2+1)}\right]^{-1}\,.
\end{equation}
For $p_0\gg 1$ Eq.~(\ref{EnergyContent0}) becomes
\begin{equation}\label{w0_maxw-1}
  W_0 = \int\limits_0^\infty p^2\mathcal{E}f^I_0(p)dp =
  N\left[(3T-1)+ {K_1(T^{-1})\over K_2(T^{-1})}\right]\,,
\end{equation}
or for non-relativistic values of $T \ll 1$
\begin{equation}\label{w0_maxw_nonr}
  W_0 = \frac{3}{2}NT +\frac{15}{8}NT^2 + \dots
\end{equation}

Now we have (recall Eqs.~(\ref{wdot_3-1}) and (\ref{w0_dot}))
\begin{eqnarray}\label{w0_notr-1}
  &&\frac{\partial W_0}{\partial T} \frac{dT}{dt}
  = \left({W_0\over N}-\mathcal{E}_0\right)S_N
  + \int\limits_{p_0}^\infty
  \frac{p^3}{\sqrt{p^2+1}}\left[D_0(T)\frac{\partial f}{\partial p}
  -\left({dp\over dt}\right)_0 f\right]\,dp  \\
  &&=\alpha f_0 \mathcal{E}_0 p_0^{\varsigma+1}(\varsigma+1)
  \left[\frac{AQ(p_0,\varsigma)}{\alpha \mathcal{E}_0(\varsigma+1)}-1\right]
  + ATf_0 \left\{\frac{3\alpha p_0^{\varsigma+1}(\varsigma+1)}{2A}
  - \left[1+ \frac{(\varsigma+1)}{(\varsigma-1)}p_0^2\right]\right\}\,,
  \nonumber
\end{eqnarray}
where
\begin{equation}
  Q(p_0,\varsigma) = \int\limits_{p_0}^\infty x^{-\varsigma}
  \sqrt{x^2+1\,} \,dx \,.
\end{equation}
If $\alpha \mathcal{E}_0(\varsigma+1) \neq AQ(p_0,\varsigma)$ and
$\mathcal{E}_0 \gg T$ then the second term in Eq.(\ref{w0_notr-1})
is small and can be neglected. Finally from
Eq.~(\ref{w0_maxw_nonr}) we obtain
\begin{equation}\label{tdot_rel-1}
  \frac{dT}{dt} =
  \frac{2S_N}{3N}\left[\frac{A Q(p_0,\varsigma)}{\alpha(\varsigma+1)} -
 \mathcal{E}_0 \right]\,,
\end{equation}
where $A$ and $S_N$ are defined by Eqs. (\ref{taumc-1}) and
(\ref{SN2-1}).

For high values of $\alpha$ one can see from Eq.~(\ref{tdot_rel-1})
that the plasma cools down and $dT/dt < 0$. The temperature
decreases with time  due to a very intensive  outflow of high
energy particles from the thermal pool, even though external sources in
the form of stochastic Fermi acceleration supply energy to the system
(analogue to Maxwell demon). This effect can be seen in Fig.
\ref{fig:spectr} as a deficit of high energy thermal particles at
$p<p_0$.

When $\alpha$ decreases, collisions start to dominate over the
outflow effect. As the result the derivative $dT/dt$ increases and
at sufficiently small $\alpha$ the regime changes from cooling to
heating of plasma. However the process of acceleration reduces the
amount of particles in the thermal pool ($dN/dt<0$ at $p<p_0$). If
$\alpha$ remains constant then the value of $A$ decreases with
time and, in principle for a sufficiently long time we come again to the
condition when $\alpha \mathcal{E}_0(\varsigma+1) >
A(t)Q(p_0,\varsigma)$, enter the regime of plasma cooling again.

A more accurate analysis of this regime can be provided by
numerical calculations of the nonlinear case.

\section{Nonlinear Case: Semi-Analytical Method and Numerical Calculations}\label{nl}

The most straightforward way to solve the problem is a numerical
solution of the original nonlinear equation.
However, this method is very time-consuming. We proceed with approximation methods
that simplify the numerical calculations, but still give a good result.

\begin{figure}[ht]
\begin{center}
\includegraphics[width=0.6\textwidth]{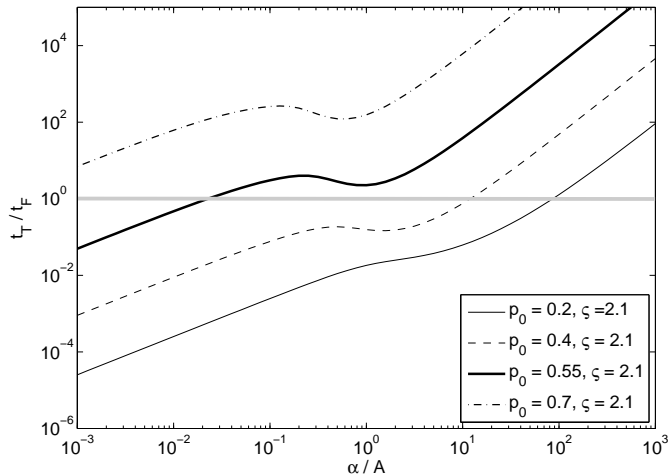}
\end{center}
\caption{The comparison between heating timescale $t_T$ and
tail-formation time-scale $t_{\rm F}$ for different $p_0$ and
acceleration rates. The temperature is $T=0.016$ (the corresponding
momentum is $p_T = 0.12$). The threshold value is marked by the gray
horizontal line.}
\label{fig:fill}
\end{figure}

Analysis of kinetic equations depend on the relation between the
plasma heating time and the acceleration time. We define the heating time as
\begin{equation}\label{t_T}
  t_T = T / (dT/dt)\,.
\end{equation}
The lower limit of this time can be obtained for the
quasi-stationary solution for $dW/dt$ when we neglect the cooling term
$S_N\mathcal{E}_0$ in Eq.~(\ref{wdot_3-1}). The acceleration time
characterizes a period  required for particles to fill the
non-thermal tail. Numerical calculations show that for
$\varsigma>2$ this time is of the order of
\begin{equation}\label{t_acc}
  t_{\rm F} \simeq \alpha^{-1} \,.
\end{equation}
The quasi-stationary state (when the plasma temperature is almost
constant and the acceleration generate prominent non-thermal
``tails'') can be reached only if $t_T > t_{\rm F}$.
In this case we can use analytical solutions presented in previous section.
The ratio $t_T/t_{\rm F}$ as a function of $p_0$ is
shown in Fig.~\ref{fig:fill}. The threshold value of ratio
$t_T/t_{\rm F} = 1$ is shown in Fig.~\ref{fig:fill} by the gray
horizontal dashed line. The quasi-stationary state will be achieved if
$t_T/t_{\rm F}$ is above the gray line.

If the acceleration time is larger than the heating time, $t_T < t_{\rm F}$,
the quasi-stationary state cannot be reached. In this case we can
simplify the calculations using the trick in \cite{east}. The
evolution of distribution function $f(p)$ can be described by the
non-stationary linear kinetic equation
\begin{equation}
  \frac{\partial f}{\partial t} +
  \frac{1}{p^2}\frac{\partial}{\partial p}p^2
  \left[ \left({dp\over dt}\right)_0(p,N,T)f \label{eq_lin_num_0}
  -\left\{D_0(p,N,T)+D_F(p)\right\}\frac{\partial f}{\partial p}
  \right] = 0\,.
\end{equation}
We can estimate the variation of temperature by the following algorithm:
\begin{enumerate}
  \item For a given $f(t,p)$, estimate $f(t+\delta t, p)$ from
        Eq.~(\ref{eq_lin_num_0});
  \item compute $N(t+\delta t)$ from $\int_0^\infty f(t+\delta t,p) dp$;
  \item calculate $\dot{W_0}$ from Eq.~(\ref{wdot_3-1}),
        then $W_0(t+\delta t) = W_0(t)+\dot{W_0} \delta t$,
        then find $T(t+\delta t)$ from Eq.~(\ref{EnergyContent0});
  \item for new values of $N(t+\delta t)$ and $T(t+\delta t)$ recalculate the
        kinetic coefficients using analytical expressions for Maxwellian
        coefficients, see Eqs.~(\ref{MaxwellCoeff1}) \& (\ref{MaxwellCoeff2});
  \item repeat steps 1-4.
\end{enumerate}

The analytical expressions for the kinetic coefficients are
calculated as
\citep[see][and references therein]{east}:
\begin{equation}\label{MaxwellCoeff1}
  \left({dp\over dt}\right)_0(p,N,T)
  = -\,\frac{A(p^2+1)}{p^2}\left[\mbox{erf}\left(\sqrt{{\mathcal{E}\over T}}
  \right) - \sqrt{{4\mathcal{E}\over \pi T}}\exp\left(-{\mathcal{E}\over T}
  \right)\right]\,,
\end{equation}
\begin{equation}\label{MaxwellCoeff2}
  D_0(p,N,T) = -\, {T\sqrt{p^2+1}\over p}\left({dp\over dt}\right)_0(p,N,T) \,.
\end{equation}
Here $\mbox{erf}(z)$ is the error function.

With this method we combine the simplicity of the analytical method
with the accuracy of the numerical method. The only problem is that
this approach like any other semi-analytical method based on Eq.
(\ref{wdot_3-1}) cannot be used near $p_0 = 0$.

Now we compare the results obtained with different methods. We
consider the following methods:
\begin{enumerate}
  \item Transitionless case: It is based on Eq.~(\ref{tdot_rel-1}). The equations
        are integrated numerically using the Runge-Kutta method to obtain the evolution
        of the temperature $T(t)$ and density $N(t)$.  This method is  applicable if
        $p_{\rm inj} < p_0$.
  \item Quasi-linear approximation: The distribution function is given
        as in Eq. (\ref{f_1_bigO-1}) and (\ref{gur_sol01-1}). Eqs.~(\ref{wdot_3-1})
        and (\ref{EnergyContent0}) are used to estimate the $dT/dt$.
        The variations of the temperature $T(t)$ and density $N(t)$ are obtained using
        the Runge-Kutta method. This approximation is valid for $t_T > t_{\rm F}$.
  \item Semi-analytical method: It uses a combination of numerical solution to
        Eq.~(\ref{eq_lin_num_0}) and analytical calculations of Eqs.~(\ref{wdot_3-1})
        \& (\ref{EnergyContent0}) in order to estimate the variations of the temperature.
        This method can be applied to $p_0 \gg p_T = \sqrt{(T+1)^2-1}$.
  \item Numerical method: Evolution of the distribution function is obtained by a numerical
        solution of the original non-linear equation Eq.~(\ref{e_k})
        (for details see Appendix~\ref{appendix-num}).
        This method is the most universal and is used to check whether the results obtained
        by methods 1-3 are correct.
\end{enumerate}

\begin{figure}[h]
\begin{center}
\includegraphics[width=0.6\textwidth]{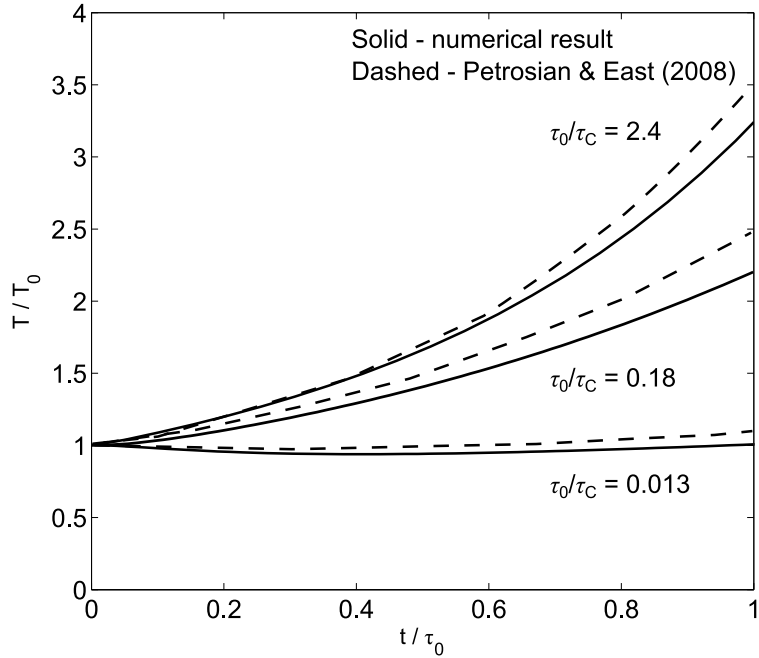}
\end{center}
\caption{Comparison between our numerical method (solid line) and
method used by \citet{east} (dashed line). All notations are the
same as in \citet{east} (see the text for details).}
\label{fig:petr_2}
\end{figure}

\begin{figure}[h]
\begin{center}
\includegraphics[width=0.6\textwidth]{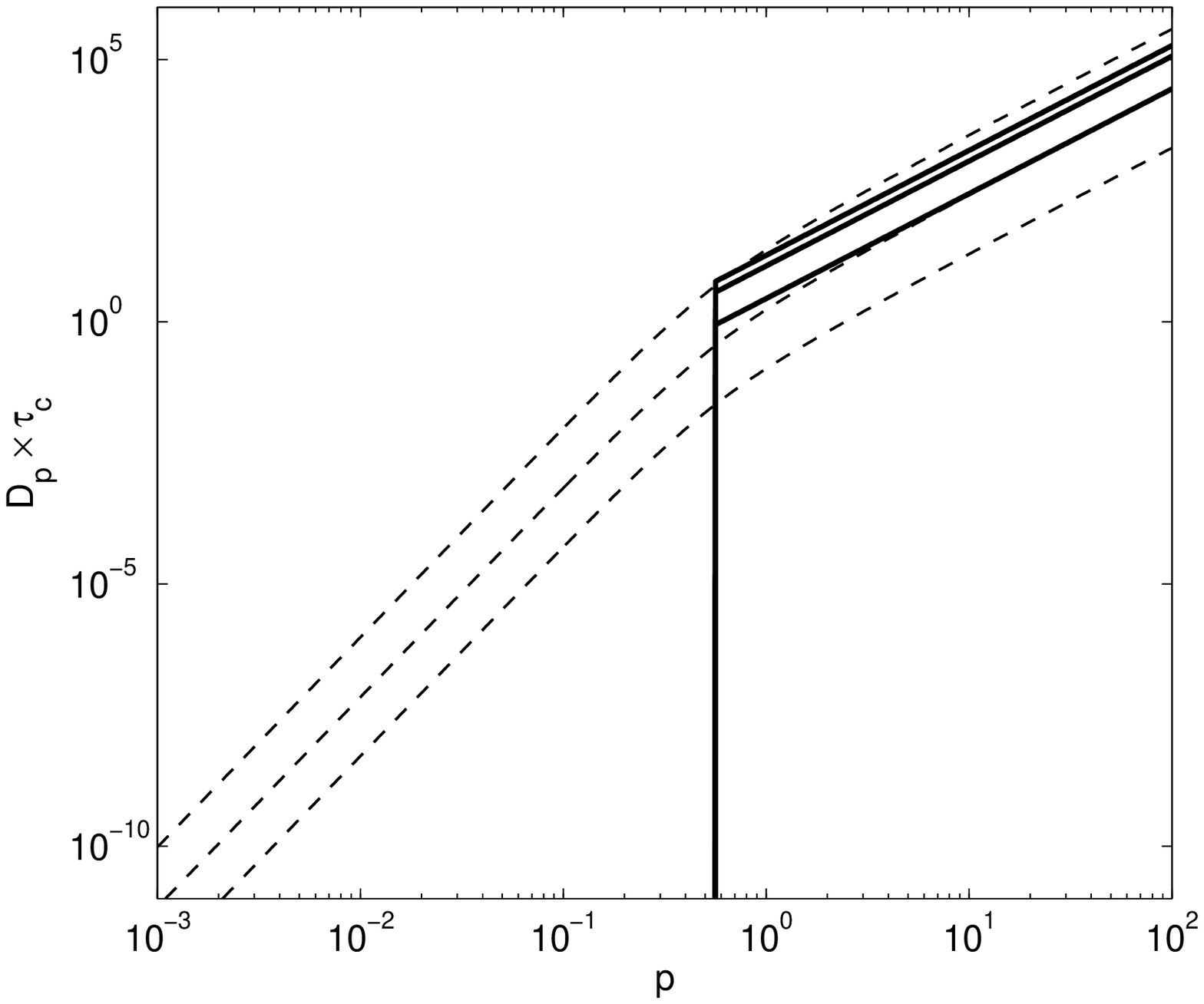}
\end{center}
\caption{Comparison between $D_F(p)$ used by \citet{east} (thin
dashed lines), Eq.~(\ref{petrosian}), and in this paper (thick
solid lines), Eq.~(\ref{pcut}).}
\label{fig:Dp}
\end{figure}

We checked our numerical program by calculations of temperature
variations for the acceleration in the form Eq.~(\ref{petrosian})
and for the same parameters  as used by \citet{east}, i.e.,
$p_0 = 0$, $q = 1$, $\mathcal{E}_c = 0.2$ and three values of $\tau_0$
equaled correspondingly: $2.4\tau_{C}$, $0.18\tau_{C}$ and
$0.013\tau_{C}$, where $\tau_{C} \equiv (4\pi r_0^2 c N \ln
\Lambda)^{-1} \approx 2.7\times 10^7\times
(N/10^{-3}~\mbox{cm}^{-3})^{-1}$ yr and $r_0 = e^2/(m_ec^2)$. The
acceleration parameters, $D_F(p)$, for these three cases are shown
in Fig. \ref{fig:Dp} by the thin dashed lines.

The result of our calculations and that of \citet{east} are shown
in Fig.~\ref{fig:petr_2} by the solid and dashed lines
correspondingly. One can see that despite of some discrepancy the
results are more or less the same.

Now we present results of calculations for the acceleration
parameter $D_F$ in the form (\ref{pcut}), when $p_0\neq 0$. In all
cases we take $p_0=0.55$.
Variations of $p_0$ change the temperature variations
quantitatively but not qualitatively.
We choose $\varsigma = 2$ in order to
obtain the same momentum dependence of $D_F$ at high energies  as
in Eq.~(\ref{petrosian}). Note that in this case $t_{\rm F} = 2\tau_0$.
Below we will use $\tau_0$ as a characteristic timescale to compare results
with those of \citet{east}.

For other values of $\varsigma$ the results are qualitatively the same, yet
lower values of $\varsigma$ will increase the amount of non-thermal particles
and thus decrease the heating timescale and vice-versa. One can see this from
Eq. (\ref{tdot_rel-1}).

The calculations where performed for the three different regimes
of acceleration:
\begin{itemize}
\item[(a)] heating dominates over cooling;
\item[(b)] cooling and heating rates are of the same order;
\item[(c)] cooling dominates over heating.
\end{itemize}
The functions $D_F(p)$ used for these three cases are shown in
Fig.~\ref{fig:Dp} by the solid lines.

We provide calculations by the four different methods: analytical
(transitionless), quasi-linear, semi-analytical and numerical.
Temperature variations, $T(t)$, obtained by these methods are
shown in Figs. \ref{fig:Tneff}, \ref{fig:Teff} and \ref{fig:Ttless}
by the dashed, thin solid, thick solid and dotted lines, correspondingly.
\begin{itemize}
\item [(a)] For the case of slow acceleration  we take $\alpha/A =
2.77$. In this case the quasi-stationary approach is not valid,
and only numerical and semi-analytical methods can provide an
adequate result. Temperature variations for this case of
acceleration parameter are shown in Fig.~\ref{fig:Tneff}. As one
can see  the result of acceleration for this parameter $\alpha$ is
the plasma overheating that is in
complete agreement with the conclusions of \citet{wolfe} and
\citet{east}.
The only difference is that the overheating occurs for the time $t
\approx 4.5\tau_0 \approx 1.6\tau_C$ which is longer than that of
\citet{east} by a factor of 4 who obtained $t\sim \tau_0$. The
reason is that for $p_0 = 0$  the acceleration generates an
extended excess above the equilibrium Maxwellian function while
for $p_0 \neq 0$ this excess is not so prominent (compare dashed and solid lines in
Fig.~\ref{fig:trans_petr}). Since the amount of suprathermal
particles in the case $p_0=0$ is higher than the case $p_0>0$,
it is not surprising that the plasma is overheated by the Coulomb
losses in a shorter time when $p_0=0$.

\begin{figure}[ht]
\begin{center}
\includegraphics[width=0.6\textwidth]{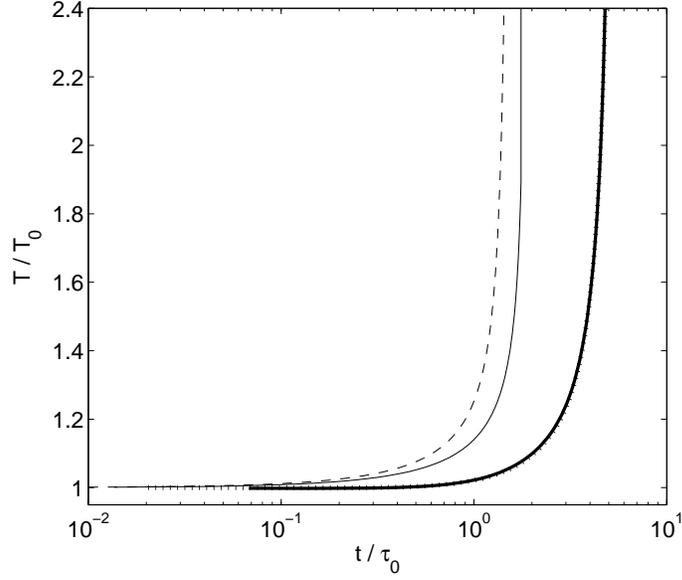}
\end{center}
\caption{The temperature evolution when heating dominates over
cooling for the parameters: $\alpha/A = 2.77$, $p_0 = 0.55$,
$\varsigma = 2$, $T_0 = 7$ keV. Dotted line - numerical model,
thick solid line - semi-analytical model, thin solid line -
quasi-linear model, dashed line - transitionless model. Note that
semi-analytical model and the numerical model almost overlap entirely.}
\label{fig:Tneff}
\end{figure}

\begin{figure}[ht]
\begin{center}
\includegraphics[width=0.6\textwidth]{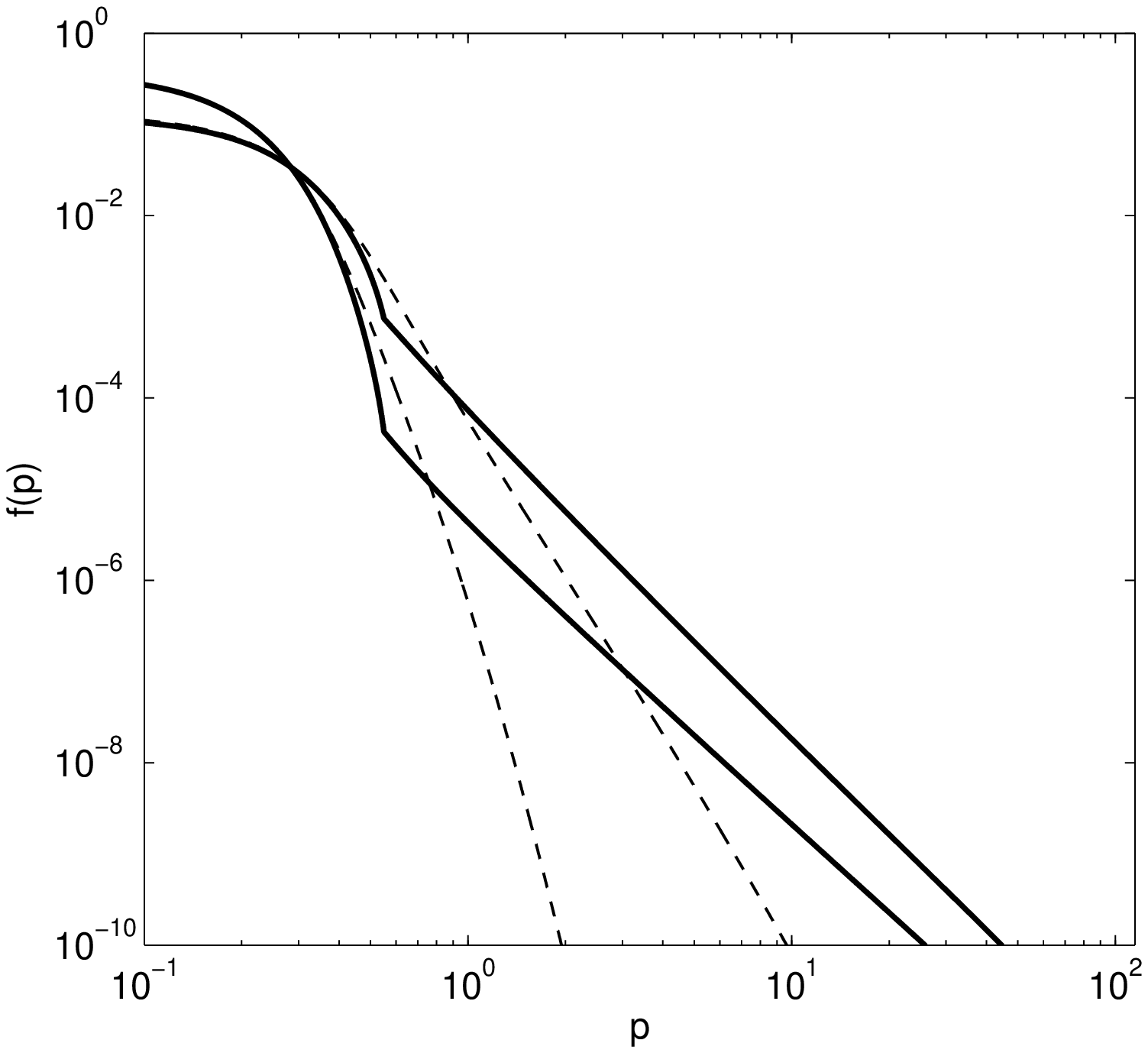}
\end{center}
\caption{Comparison between spectra formed under influence of
momentum diffusion coefficient in the form of
Eq.~(\ref{petrosian}) for $\tau_0 = 0.18\tau_{C}$ (thin dashed
lines) and those of Eq.~(\ref{pcut}) for $\alpha/A=2.77$,
$\varsigma=2$ and $p_0=0.55$ (solid lines). The distribution
were taken at the moment when the temperatures are the same.
Two values of the temperature were used: $T_1 = 8.5$ keV,
$T_2 = 16.5$ keV.}
\label{fig:trans_petr}
\end{figure}

\item [(b)] The case of moderate acceleration ($\alpha/A = 11.63$) is
shown in Fig.~\ref{fig:Teff}.
One can see that all methods are in good agreement. At the first
stage we see plasma heating,  however the timescale is much longer
than in \citet{east}, the plasma temperature  increases by a
factor of 1.3 at the moment $t  \approx 86\tau_0 = 7\tau_C$ that
is almost two orders of magnitude higher than that of
\citet{east}. A prominent quasi-stationary power-law
tail of nonthermal particles is formed by the acceleration
for a much shorter time (since $t_T/t_{\rm F} > 1$, see Fig. \ref{fig:fill}).
Moreover unlike in \citet{east}, after this time heating reverses to cooling.
\item [(c)] The case of fast acceleration
($\alpha/A = 18.5$) is shown in Fig. \ref{fig:Ttless}.
The numerical method, the semi-analytical method and the quasi-linear
approximation give almost the same result. For comparison we also
show the calculations obtained with transitionless case (dashed
line). We see that in spite of some difference this method
provides a similar qualitative time variations of the temperature
$T$. All methods demonstrate plasma cooling in this regime from
the very beginning. The temperature of the plasma shows a steady
decrease with time while nonthermal tails are formed rapidly ($t_T/t_{\rm F} > 1$,
see Fig. \ref{fig:fill}) that differs completely from the results
obtained by \citet{east}.
\end{itemize}

\begin{figure}[ht]
\begin{center}
\includegraphics[width=0.6\textwidth]{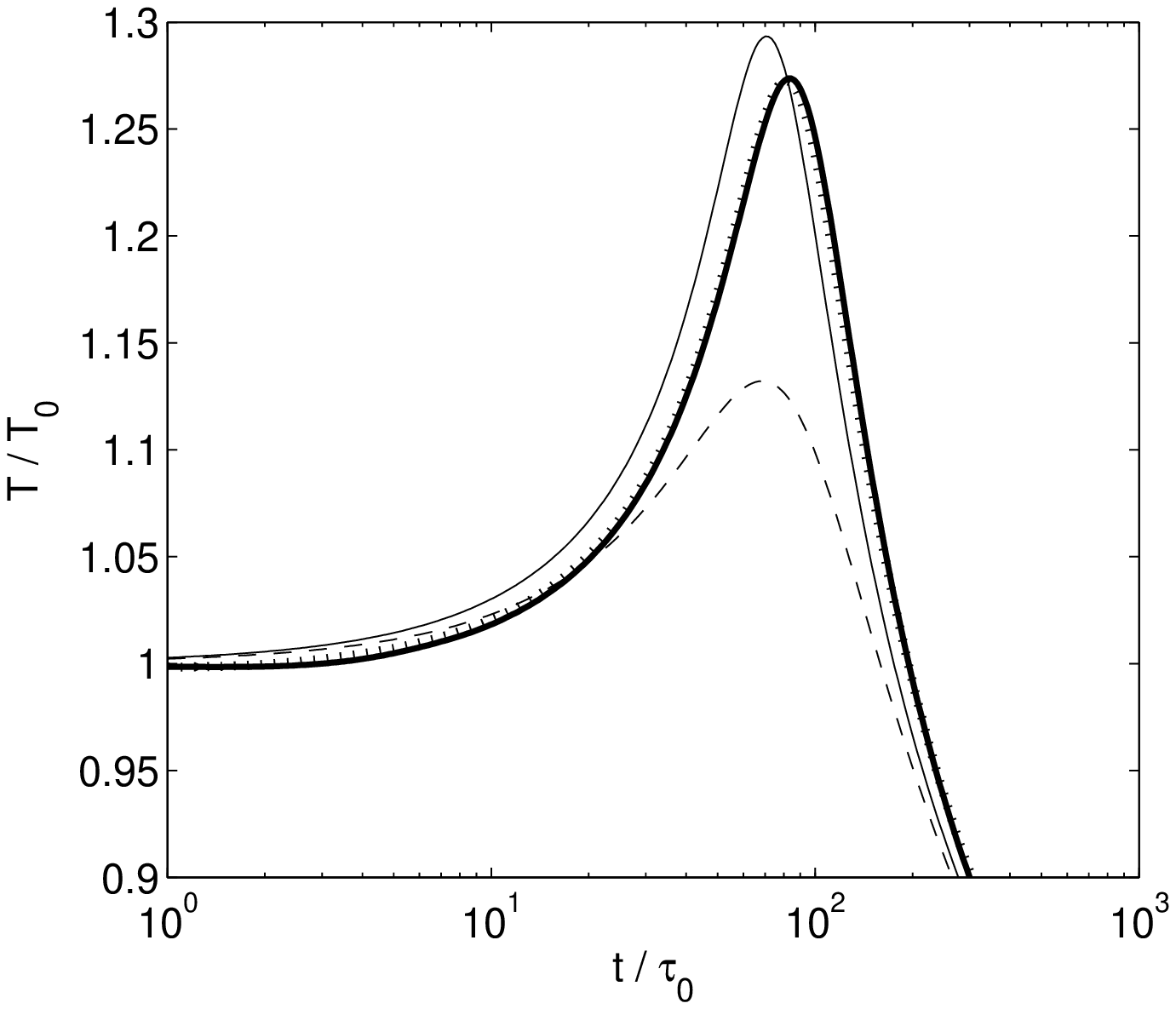}
\end{center}
\caption{Same as Fig. \ref{fig:Tneff} but for the case when
cooling and heating processes are of the same order of magnitude
($\alpha/A = 11.6$).}
\label{fig:Teff}
\end{figure}

\begin{figure}[ht]
\begin{center}
\includegraphics[width=0.6\textwidth]{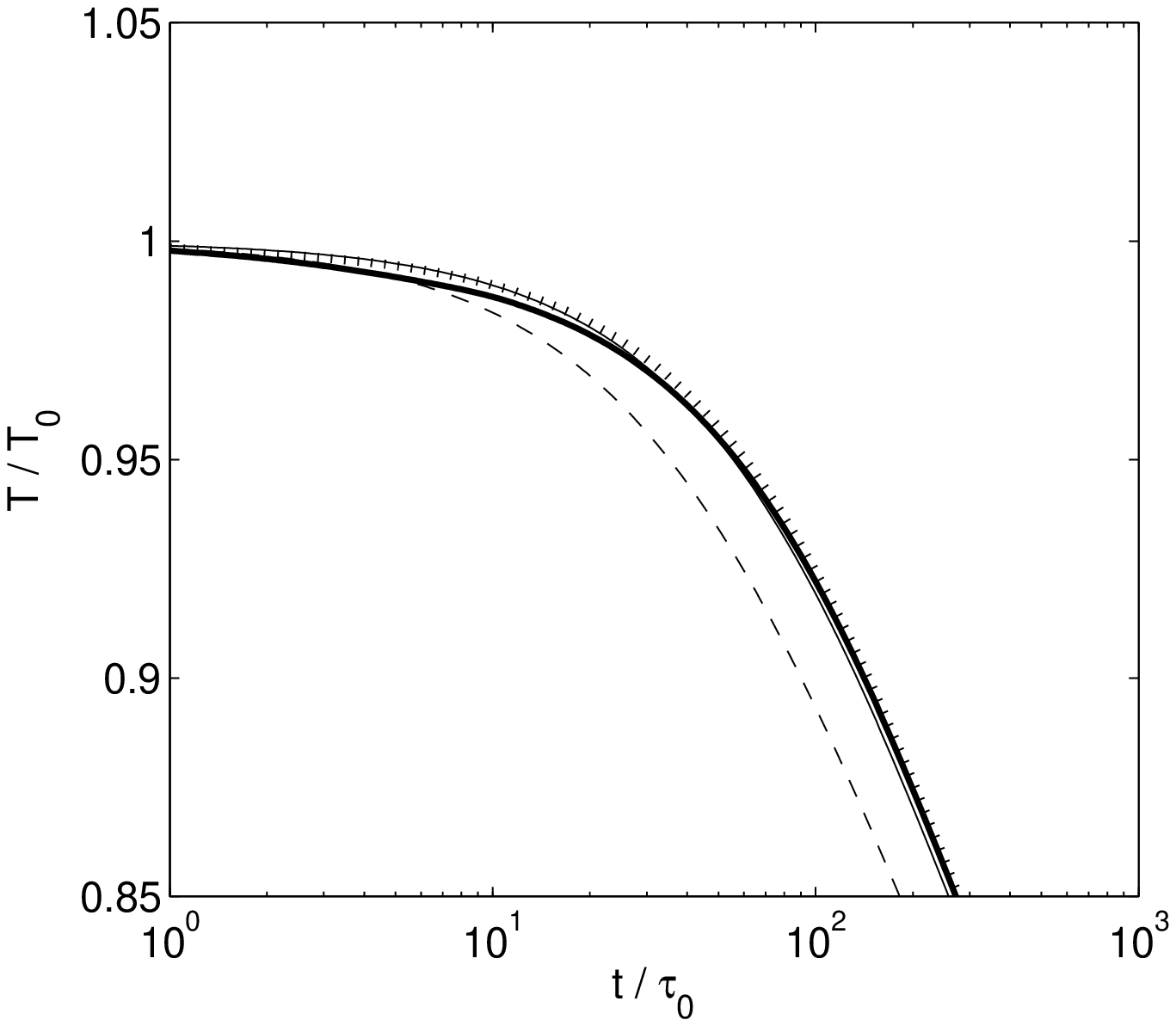}
\end{center}
\caption{Same as Fig. \ref{fig:Tneff} but for the case when
cooling dominates over heating ($\alpha/A = 18.5$).}
\label{fig:Ttless}
\end{figure}

These results can be understood from Fig. \ref{fig:spectr}.
For the ratio $\alpha/A = 2.77$ the injection momentum is
$p_{\rm inj}\simeq 0.83$, i.e., $p_{\rm inj}>p_0$
(similar to the upper curve of Fig.~\ref{fig:spectr}).
An excess of quasi-thermal particles
is formed in the range between $p_0$ and $p_{\rm inj}$.
Coulomb losses of these particles results in effective
plasma heating. As we  already above-mentioned \citet{east}
assumed $p_0=0$ that led to more extended transition region and
more effective heating.

In the case of $\alpha/A = 11.63$, $p_{\rm inj}\simeq 0.5\simeq p_0$
(similar to the middle curve of Fig.~\ref{fig:spectr}).
The transition region is almost negligible in this case.
Therefore, plasma heating by nonthermal particles is insignificant
which then changes into cooling.

In the case of $\alpha/A = 18.5$, $p_{\rm inj}\simeq 0.4$, i.e.,
$p_{\rm inj}<p_0$ (similar to the lower curve of Fig.~\ref{fig:spectr}).
A deficit of high energy particles is formed in
the thermal energy range that provides the effect of cooling.

Thus, we conclude, that depending on  parameters, $p_0$ and
$\alpha$, different regimes of acceleration from background plasma
are realized. The important inference is that stochastic
acceleration may produce a flux of nonthermal particle without
plasma overheating.

A specific spectrum of turbulence that provides stochastic
acceleration is out of the scope of this paper. It depends on
mechanisms which excite  electromagnetic fluctuations in an
astrophysical plasma. As an example, we mention particle
acceleration in OB-associations by a supersonic turbulence
\citep[see][]{byk93}. The momentum diffusion coefficient in this
case has the form $D(p)=D_0p^2$. This acceleration is effective in
the momentum range $p>p_0$, where the value of $p_0$ is derived
from $r_L(p_0)=lu/c$. Here $r_L$ is the particle
Larmor radius, $u$ is the shock velocity and $l$ is a distance
between shocks. Particles with $p<p_0$ are not accelerated by this
mechanism.

\section{Conclusion}\label{conclusion}

We analyzed nonlinear kinetic equations describing particle
stochastic (or second-order Fermi) acceleration from background
plasma when the acceleration is non-zero for particles with
momenta $p>p_0$. The goal of these investigations is to define
whether the only result of stochastic acceleration is plasma
overheating as concluded by \citet{wolfe} and \citet{east}, or this
acceleration can generate prominent tails of nonthermal particles
when the plasma temperature remains almost stationary. The
following results are obtained from our analysis:

\begin{figure}[t!]
\begin{center}
\includegraphics[width=0.6\textwidth]{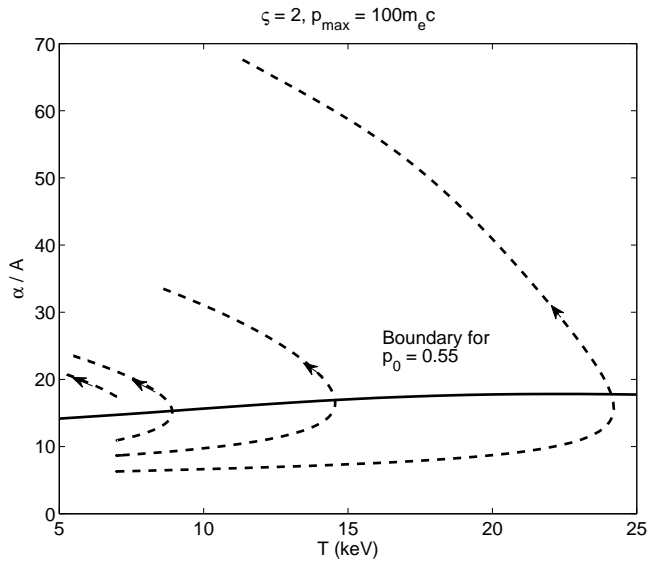}
\end{center}
\caption{Regions of heating (below the solid line) and cooling
(above the solid line) in the parameter space ($T,\alpha/A$).
The dashed lines show the evolution of systems at
the same starting temperature but with different $\alpha/A$.}
\label{traj}
\end{figure}

\begin{enumerate}
\item We showed that in the case of stochastic acceleration two
      competitive processes determine temperature variations of
      background plasma. The first one is Coulomb energy losses of
      nonthermal particles which heat the plasma. The other one is a
      run-away flux of high-energy particles from the thermal pool
      that leads to plasma cooling. Depending on the
      rates of these processes the plasma may cool down or
      heat up.
\item From numerical and analytical calculations we conclude that
      for a low enough acceleration rate the cooling process is
      negligible. The plasma gains much heat on the acceleration
      timescale $\tau_0$.
      As a result the plasma temperature rises
      rapidly while prominent nonthermal tails are not generated, that
      fully confirms results of \citet{wolfe} and \citet{east}.
\item For a moderate acceleration the cooling and heating
      processes partly compensate each other. As a results the plasma
      temperature is quasi-stationary on a timescales much longer than
      $\tau_0$.
      In this case, the acceleration produces a nonthermal
      component of the spectrum. After a period of moderate plasma
      heating the process changes into cooling. This regime does not
      appear in the models of \citet{wolfe} and \citet{east}.
\item For a high rate of acceleration the run-away flux of thermal
      particles cools the plasma down from the very beginning. In spite
      of energy supply by external sources the  plasma temperature drops
      down (analogue to Maxwell demon).
\item The  evolution of plasma temperature depends on the characteristic
      time of Coulomb collisions in the background plasma (the collision
      frequency $A$) and the acceleration frequency $\alpha$. This is
      illustrated in Fig. \ref{traj} where the solid line defines the
      border between heating (below the line) and cooling (above the
      line) regimes for quasi-stationary systems. It corresponds to the
      solution of the $dT/dt = 0$ (see Eq.~(\ref{wdot_3-1})). Dashed
      lines in Fig. \ref{traj} show the evolution of plasma parameters
      for the same initial temperature but different initial value of
      $A$ ($\alpha$ is the same for the systems). Since $N$ decreases
      monotonically because of particle acceleration, $A$ decreases
      monotonically accordingly (see Eq.~(\ref{taumc-1})).

      One can see that even if the system starts from the
      regime of heating,  sooner or later it changes to plasma
      cooling. If the evolution of the system is quasi-stationary the turning point of
      the trajectory should be located on the boundary. However we mention that the
      quasi-stationary approximation is inapplicable to low values of $\alpha / A$.
\end{enumerate}

\appendix

\section{General Kinetic Equation}\label{appendix-kin}

The general equation for stochastic Fermi acceleration with the
coefficient $D^F_{\alpha\beta}$ the equation has the form
\citep{ll, wolfe}
\begin{equation}\label{equp}
  \frac{\partial f({\bf p})}{\partial t}=\frac{\partial}{\partial p_\alpha}
  \left[\left(D_{\alpha\beta}+D^F_{\alpha\beta}\right)\frac{\partial f({\bf p})}
  {\partial p_\beta}-F_\alpha f({\bf p})\right]\,,
\end{equation}
where ${\bf p}= {\bf v}\left(c^2- {\bf v}^2\right)^{-1/2}$ is the
dimensionless particle momentum and ${\bf v}$ is the particle
velocity. The coefficients $D_{\alpha\beta}$ and $F_\alpha$ are
determined by Coulomb collisions of the particles,
\begin{equation}\label{coef1p}
  D_{\alpha\beta}=A\int Z_{\alpha\beta}({\bf p},{\bf
  p^\prime})f({\bf p^\prime})d^3 p^\prime\,,
  \quad F_\alpha=-A\int\left[\frac{\partial}{\partial p^\prime_\beta}
  Z_{\alpha\beta}({\bf p},{\bf p^\prime})\right]f({\bf p^\prime})d^3 p^\prime\,,
\end{equation}
where
\begin{equation}\label{coef2p}
  Z_{\alpha\beta}({\bf p},{\bf p^\prime})=\frac{r^2}{\gamma\gamma^\prime w^3}
  \left[w^2\delta_{\alpha\beta}-p_\alpha p_\beta-p^\prime_\alpha p^\prime_\beta
  +r(p_\alpha p^\prime_\beta+p^\prime_\alpha p_\beta)\right]\,,
\end{equation}
\begin{equation}\label{coef3p}
  A=\frac{8\pi e^2e^{\prime\,2}\ln\Lambda}{m^2}\,,
  \quad r=\gamma\gamma^\prime-{\bf p\cdot p^\prime}/c^2\,,
  \quad w=c\sqrt{r^2-1}\,,
  \quad \gamma=\sqrt{1+{\bf p}^2/c^2}\,.
\end{equation}
The boundary conditions were taken in the form: a zero particle
flux at $p=0$:
\begin{equation}\label{kinet_symp}
  \left[\left(D_{\alpha\beta}+D^F_{\alpha\beta}\right)
  \frac{\partial f({\bf p})}{\partial p_\beta}-F_\alpha f({\bf p})\right]_{p=0} = 0\,,
\end{equation}
and the distribution function vanishes at some $p_{\rm max}$:
\begin{equation}\label{kinet_sym1p}
  f({\bf p}_{\rm max}) = 0\,.
\end{equation}

\section{Numerical Method for the Nonlinear Case}\label{appendix-num}

To solve Eq.~(\ref{e_k}) numerically we use the
Crank-Nicolson finite difference method. To estimate the kinetic
coefficients~(\ref{coef1p}) we use Simpson's integration rule,
\begin{equation}\label{discr_coef}
  {\bf D} = {\bf \mathcal{Z}f}\,,\quad {\bf F}
  = {\bf \mathcal{Z^\prime}f}\,,
\end{equation}
where vectors ${\bf f}$, ${\bf D}$ and ${\bf F}$ are
corresponding discrete versions of $f(p)$, $D_c(p,f)$ and
$(dp/dt)_c(p,f)$. Matrices ${\bf\mathcal{Z}}$ and
${\bf\mathcal{Z^\prime}}$ are obtained by applying Simpson's
rule to Eq.~(\ref{coef1p}). The discrete version of Eq.~(\ref{e_k})
at $t = (t_n+t_{n+1})/2$ and $p = p_j$ looks like (see e.g., \cite{park} and
references therein)
\begin{equation}\label{discreet1}
  \frac{f_{n+1,j} - f_{n,j}}{\Delta t} =
  \frac{1}{p_j^2}\frac{S_{n+{\scriptstyle{1\over2}},j+{\scriptstyle{1\over2}}}
  -S_{n+{\scriptstyle{1\over2}},j-{\scriptstyle{1\over2}}}}{\Delta p_j} \,,
\end{equation}
where $\Delta t = t_{n+1}-t_{n}$ and $\Delta p_j = (p_{j+1}-p_{j-1})/2$
are steps of the grid and the flux is expressed according to the
Crank-Nicolson rule:
\begin{eqnarray}
  S_{n+{\scriptstyle{1\over2}},j+{\scriptstyle{1\over2}}}
  &&= {\textstyle{1\over2}} p_j^2\left[D_{n+1,j+{\scriptstyle{1\over2}}}
  \frac{f_{n+1,j+1}-f_{n+1,j}}{\Delta p_{j+{\scriptstyle{1\over2}}}}
  - F_{n+1,j+{\scriptstyle{1\over2}}}f_{n+1,j+{\scriptstyle{1\over2}}}\right] + \\
  && \quad + {\textstyle{1\over2}} p_j^2
  \left[D_{n,j+{\scriptstyle{1\over2}}}\frac{f_{n,j+1}-f_{n,j}}
  {\Delta p_{j+{\scriptstyle{1\over2}}}}
  - F_{n,j+{\scriptstyle{1\over2}}}f_{n,j+{\scriptstyle{1\over2}}}\right] \nonumber \,,\\
  D_{n,j+{\scriptstyle{1\over2}}} &&= {\textstyle{1\over2}}(D_{n,j}+D_{n,j+1}) \,, \\
  F_{n,j+{\scriptstyle{1\over2}}} &&= {\textstyle{1\over2}}(F_{n,j}+F_{n,j+1}) \,, \\
  f_{n,j+{\scriptstyle{1\over2}}} &&= {\textstyle{1\over2}}(f_{n,j}+f_{n,j+1}) \,, \\
  \Delta p_{j+{\scriptstyle{1\over2}}} &&= p_{j+1}-p_j \,.
\end{eqnarray}
The boundary condition at $p=0$ imply that
$S_{n+{\scriptstyle{1\over2}},-{\scriptstyle{1\over2}}} = 0$.
The boundary condition at $p=p_{\rm max}$ is
$f(p_{\rm max})=0$.

After discretization we arrive at the non-linear system of equations,
\begin{equation}\label{eq_dsk}
  {\bf f_{n+1}} - {\bf f_{n}} = \mathcal{A}({\bf f_{n+1}}){\bf f_{n+1}}
  + \mathcal{A}({\bf f_{n}}){\bf f_{n}} \,,
\end{equation}
where ${\bf \mathcal{A}(f)}$ is a tridiagonal matrix corresponding
to the differential operator in RHS of Eq.~(\ref{discreet1}).
According to Eq.~(\ref{discr_coef}), ${\bf \mathcal{A}(f)}$ is a
linear function of ${\bf f}$ and Eq.~(\ref{eq_dsk}) is a system of
quadratic equations.

To avoid calculations of Jacobian matrix we do not apply Newton's
method and utilize a simple iteration method instead. However the
iteration method based on Eq.~(\ref{eq_dsk}) converges very slowly.
We rewrite the iteration step in the following form
\begin{equation}\label{eq_dsk2}
  {\bf \left(E-\mathcal{A}(f^k_{n+1})\right)f^{k+1}_{n+1}} =
  \mathcal{A}({\bf f_{n}}){\bf f_{n}} + {\bf f_{n}}\,,
\end{equation}
where $k$ is the number of iteration and ${\bf E}$ is a unit matrix.
The system of linear equations is solved using tridiagonal matrix algorithm.
Iteration in the form Eq.~(\ref{eq_dsk2}) shows fast convergence if the
temperature of the Maxwellian distribution does not change significantly
between $t_n$ and $t_{n+1}$.

Since the Crank-Nicolson method may be affected by numerical oscillations we
also use less precise and more robust backward Euler method
\citep[simple fully implicit method from][]{park}.
The backward Euler method turns out to be
useful for non-thermal tails of low magnitude when the value of $\alpha$ is low
and the value of $p_0$ is high.

The discretization in momentum space is tricky since we need to
provide a good resolution for Maxwellian distribution and
transitional region as well as calculate the nonthermal tail at
high energies. We tried two possible ways to reduce the number of
grid points. The first one is to split the momentum axis
into sub-domains and join them using continuity of the
distribution function and particle flux. The second way is to use
the logarithmic grid by introducing new variable $q=\log(p)$.
Both methods gives almost the same results.

\section*{Acknowledgements}

We thank the anonymous referee for valuable comments on an earlier
version of the paper. DOC and VAD are partly supported by the RFFI
grant 12-02-00005-a.  CMK is supported, in part, by the Taiwan
National Science Council under the grants NSC 98-2923-M-008-01-MY3
and NSC 99-2112-M-008-015-MY3.

\end{document}